\documentclass[prl,twocolumn,amsfonts,superscriptaddress,showpacs]{revtex4}
 
\usepackage[]{graphicx} 
\usepackage{amsbsy} 
\usepackage{float}
\usepackage{color}
\usepackage{subfigure}
 
\newcommand{\be}{\begin{equation}}
\newcommand{\ee}{\end{equation}}
\newcommand{\bea}{\begin{eqnarray}}
\newcommand{\eea}{\end{eqnarray}}
\newcommand{\ba}{\begin{eqnarray*}}
\newcommand{\ea}{\end{eqnarray*}}

\begin{document} 
\title{Ultrafast evolution and transient phases of a prototype out-of-equilibrium Mott-Hubbard material}

\author{G.~Lantz} 
\affiliation{Laboratoire de Physique des Solides, CNRS, Univ. Paris-Sud, Universit\'{e} Paris-Saclay, 91405 Orsay, France}
\affiliation{Institute for Quantum Electronics, Physics Department, ETH Zurich, CH-8093 Zurich, Switzerland}
\author{B.~Mansart} 
\affiliation{Laboratoire de Physique des Solides, CNRS, Univ. Paris-Sud, Universit\'{e} Paris-Saclay, 91405 Orsay, France}
\author{D.~Grieger} 
\affiliation{International School for Advanced Studies SISSA, Via Bonomea 265, 34136 Trieste, Italy}
\author{D.~Boschetto}
\affiliation{Laboratoire d'Optique Appliqu\'{e}e, ENSTA, CNRS, Ecole Polytechnique, F-91761 Palaiseau, France}
\author{N.~Nilforoushan} 
\affiliation{Laboratoire de Physique des Solides, CNRS, Univ. Paris-Sud, Universit\'{e} Paris-Saclay, 91405 Orsay, France}
\author{E.~Papalazarou} 
\affiliation{Laboratoire de Physique des Solides, CNRS, Univ. Paris-Sud, Universit\'{e} Paris-Saclay, 91405 Orsay, France} 
\author{N.~Moisan} 
\affiliation{Laboratoire de Physique des Solides, CNRS, Univ. Paris-Sud, Universit\'{e} Paris-Saclay, 91405 Orsay, France}
\author{L.~Perfetti}
\affiliation{Laboratoire des Solides Irradi\'{e}s, Ecole Polytechnique-CEA/SSM-CNRS UMR 7642, 91128 Palaiseau, France}
\author{V.~L.~R.~Jacques}
\affiliation{Laboratoire de Physique des Solides, CNRS, Univ. Paris-Sud, Universit\'{e} Paris-Saclay, 91405 Orsay, France}
\author{D.~Le~Bolloc’h}
\affiliation{Laboratoire de Physique des Solides, CNRS, Univ. Paris-Sud, Universit\'{e} Paris-Saclay, 91405 Orsay, France}
\author{C.~Laulh\'{e}}
\affiliation{Synchrotron SOLEIL, Saint-Aubin, BP 48, 91192 Gif-sur-Yvette Cedex, France}
\affiliation{Univ. Paris-Sud, Universit\'{e} Paris-Saclay, 91405 Orsay, France}
\author{S.~Ravy}
\affiliation{Synchrotron SOLEIL, Saint-Aubin, BP 48, 91192 Gif-sur-Yvette Cedex, France}
\affiliation{Laboratoire de Physique des Solides, CNRS, Univ. Paris-Sud, Universit\'{e} Paris-Saclay, 91405 Orsay, France}
\author{J.-P.~Rueff}
\affiliation{Synchrotron SOLEIL, Saint-Aubin, BP 48, 91192 Gif-sur-Yvette Cedex, France}
\author{T.E.~Glover}
\affiliation{Advanced Light Source, Lawrence Berkeley National Laboratory, Berkeley, CA 94720, USA}
\author{M.P.~Hertlein}
\affiliation{Advanced Light Source, Lawrence Berkeley National Laboratory, Berkeley, CA 94720, USA}
\author{Z.~Hussain}
\affiliation{Advanced Light Source, Lawrence Berkeley National Laboratory, Berkeley, CA 94720, USA}
\author{S.~Song}
\affiliation{SLAC National Accelerator Lab, Stanford PULSE Inst, Menlo Pk, CA 94025, USA}
\author{M.~Chollet}
\affiliation{SLAC National Accelerator Lab, Stanford PULSE Inst, Menlo Pk, CA 94025, USA}
\author{M.~Fabrizio} 
\affiliation{International School for Advanced Studies SISSA, Via Bonomea 265, 34136 Trieste, Italy}
\author{M.~Marsi}
\affiliation{Laboratoire de Physique des Solides, CNRS, Univ. Paris-Sud, Universit\'{e} Paris-Saclay, 91405 Orsay, France}
\date{\today}

\maketitle 

\textbf{
The study of photoexcited strongly correlated materials is attracting 
growing interest since their rich phase diagram often translates into an equally rich out-of-equilibrium behavior, including non-thermal phases and photoinduced phase transitions. With femtosecond optical pulses, electronic and lattice degrees of freedom can be transiently decoupled, giving the opportunity of stabilizing new states of matter inaccessible by quasi-adiabatic pathways. Here we present a study of the ultrafast non-equilibrium evolution of the prototype Mott-Hubbard material V$_2$O$_3$, which presents a transient non-thermal phase developing immediately after photoexcitation and lasting few picoseconds. For both the insulating and the metallic phase, the formation of the transient configuration is triggered by the excitation of electrons into the bonding $a_{1g}$ orbital, and is then stabilized by a lattice distortion characterized by a marked hardening of the  $A_{1g}$ coherent phonon. This configuration is in stark contrast with the thermally accessible ones - the $A_{1g}$ phonon frequency actually softens when heating the material. Our results show the importance of selective electron-lattice interplay for the ultrafast control of material parameters, and are of particular relevance for the optical manipulation of strongly correlated systems, whose electronic and structural properties are often strongly inter-twinned.}

The Mott metal-to-insulator transition (MIT)\cite{Mott1937} is the perfect example of how thermodynamic parameters can affect the electronic structure of a material and its conducting properties. At equilibrium, temperature, doping, and pressure act as driving forces for such transitions \cite{Imada1998}, that invariably involve also a lattice modification - either with a change of symmetry, like for instance in VO$_2$ \cite{Goodenough1971} or with a lattice parameter jump like in V$_2$O$_3$ \cite{McWhan1969}. It is actually often unclear whether the lattice or the electronic structure is the trigger for the MIT since at equilibrium both change together. This "chicken and egg" question can be answered by driving one far from equilibrium and observing the reaction of the other. Thus, time-resolved pump-probe techniques \cite{Cavalleri2001,Wegkamp2014a,Perfetti2006,Yoshida2014,Morrison2014} can provide this answer, as long as the response of the electrons and of the lattice can be separately analyzed.

\begin{figure}
\includegraphics[angle=0,width=1\linewidth,clip=true]{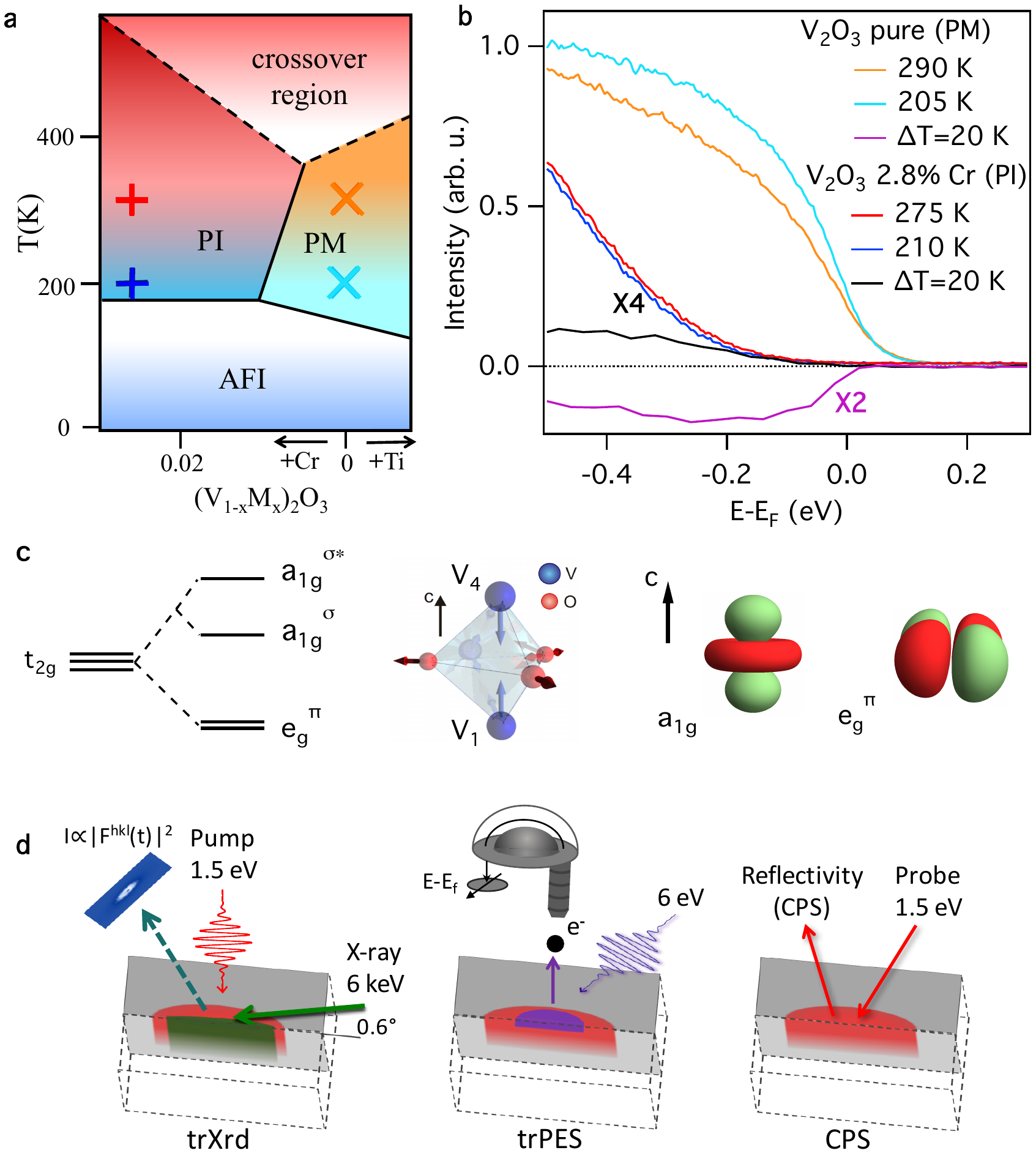}
\caption{\textbf{a)}: (V$_{1-x}$Cr$_x$)$_2$O$_3$ phase diagram; the crosses indicate the experimental data points. \textbf{b)}: Temperature dependence of the equilibrium photoemission spectra for (V$_{1-x}$Cr$_x$)$_2$O$_3$ (x=0.028 PI phase, x=0 PM phase). Temperature differences are shown for each doping, which are used as a thermal equilibrium reference in the comparison with the photoexcited spectra. Upon increasing the temperature, the spectral weight is transferred into the Mott gap in the PI phase, whereas the QP peak weakens in the PM phase. \textbf{c)}: representation of the orbital splitting and their geometry. \textbf{d)}: Schematic of the experiments using an optical pump and different probes: X-ray diffraction, photoemission, and coherent phonon spectroscopy.} 
\label{equi} 
\end{figure}

In this letter we use a combined experimental and theoretical approach to study the ultrafast evolution of the Mott-Hubbard prototype (V$_{1-x}$Cr$_x$)$_2$O$_3$\cite{Liu2011}. The phase diagram of V$_2$O$_3$ contains three phases, a paramagnetic metallic (PM) phase, a paramagnetic insulating (PI) phase, and antiferromagnetic insulator phase (AFI), see Fig. \ref{equi}. The isostructural Mott transition is between the PI and the PM phases \cite{Lupi2010}. In all our experiments, the energy of the pump pulses (1.5 eV) corresponds to the transition from $e_g^\pi$ to $a_{1g}$ orbitals. Thus optical pumping directly increases the $a_{1g}$ population while decreasing the $e_g^\pi$ one. Using time-resolved PhotoElectron Spectroscopy (trPES) we directly probe the electronic structure, while time-resolved X-Ray Diffraction (trXRD) and Coherent Phonon Spectroscopy (CPS) give access to the lattice evolution \cite{Rohwer2011,Hellmann2012,Johnson2009a,Boschetto2013}. Thanks to this multitechnique approach, we  can unambiguously disentangle the contribution of electrons and lattice to the non-equilibrium dynamics of the system. 
Furthermore, this archetypal material gives the opportunity of comparatively observing the ultrafast evolution of a Mott system starting both from the insulating and the metallic phase, whereas previous studies have generally focused only on the insulator as ground state \cite{Cavalleri2001,Wegkamp2014a,Perfetti2006,Yoshida2014,Morrison2014}. 
We find that in the PI phase the gap is instantaneously filled and a non-thermal transient state that lasts 2 ps is created. In the PM phase, the quasiparticle (QP) signal shows an immediate appreciable spectral redistribution across E$_F$ which also lasts 2 ps, once again not compatible with thermal heating. In both phases we find that the lattice conspires to stabilize the non-thermal transient electronic state. Ab-initio DFT-GGA results supplemented by simple Hartree-Fock calculations suggest that the gap filling is driven by the non-equilibrium population imbalance between the $e_g^\pi$ and $a_{1g}$ orbitals, which effectively weakens the correlation strength.

\begin{figure*}
\includegraphics[angle=0,width=0.75\linewidth,clip=true]{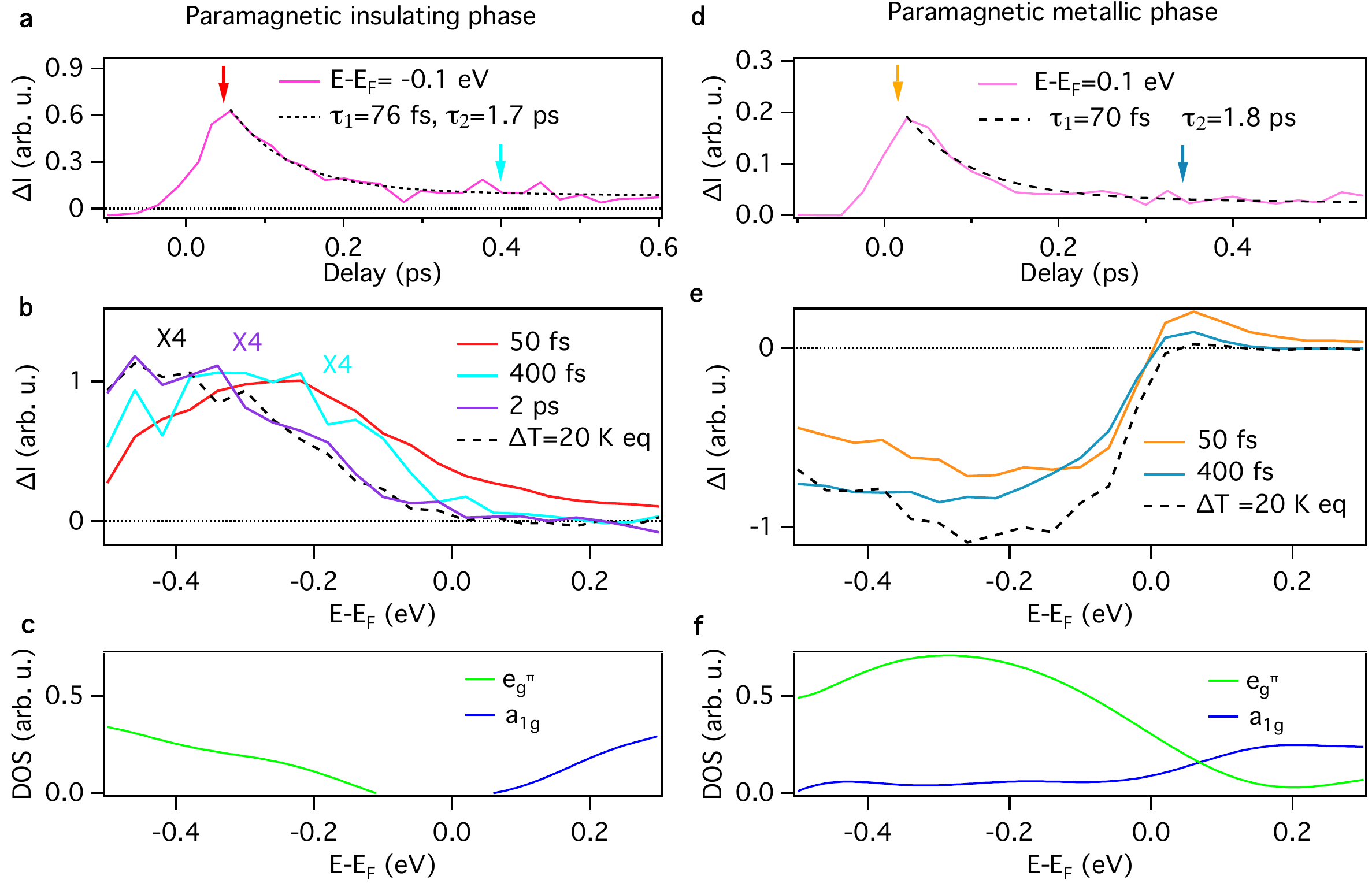}
\caption{trPES for (V$_{1-x}$Cr$_x$)$_2$O$_3$ (x=0.028 PI phase and x=0 PM phase) at a fluence of 1.8 mJ.cm$^{-2}$. \textbf{a)}: Time evolution of the intensity difference at -0.1 eV, the curve is fitted with a double exponential. \textbf{b)}: PES intensity difference for $\Delta$t=50 fs, 400 fs , and 2 ps are shown for the PI phase as well as the equilibrium temperature difference from Fig.~\ref{equi}. The 50 fs and 400 fs differences show that the spectral weight is transferred inside the Mott gap, differently from a purely thermal effect. This non-thermal distribution relaxes within 2 ps.  \textbf{c)}: Orbital character of the density-of-states near E$_F$ extracted from \citep{Poteryaev2007}. \textbf{d-f)}: Same as (\textbf{a-c}) but for the PM phase. The time evolution is fitted with a double exponential for the energy above E$_F$.}
\label{trPES} 
\end{figure*}

 In vanadium sesquioxide the octahedral crystal field leads to the $d$-orbital splitting into a lower t$_{2g}$ and an upper $e_g^\sigma$. Since the octahedron has a trigonal distortion, the t$_{2g}$ are split into a lower twofold degenerate $e_g^\pi$ orbital and an upper non-degenerate $a_{1g}$ (Fig. \ref{equi}). The hybridization between the two nearest vanadium atoms, which are lined up along the $c$-axis, causes a large splitting between bonding $a_{1g}(\sigma)$ and antibonding $a_{1g}(\sigma^*)$ states. In spite of that, the $a_{1g}$ orbital remains mostly unoccupied in the PI phase, whereas the $e_g^\pi$ orbitals are occupied by almost one electron each\cite{sah09,Poteryaev2007}. V$_2$O$_3$ PI can thus be viewed as a half-filled 2-band Mott insulator stabilized by the correlation-enhanced trigonal field that pushes above the Fermi energy (E$_F$) the $a_{1g}$ orbitals\cite{Poteryaev2007,Grieger2015}, whose occupancy indeed jumps across the doping- or temperature-driven Mott transition\cite{Park2000}, whilst is smoother across the pressure driven one\cite{Rodolakis2010,Lupi2010}. This inequivalent behavior in temperature versus pressure of the MIT and the related deep intertwining between strong correlations and lattice structure suggest that a major issue in time-resolved experiments is to distinguish a temperature increase from a transient non-thermal phase, such as hidden phases \cite{Ichikawa2011,Stojchevska2014}.

Before exploring the behavior of the system after photoexcitation, we present in Fig.~\ref{equi} the photoemission responses of the PI and PM phases at different temperatures, which give us reference energy distribution curves (EDC's) for the system at equilibrium. In the PM phase the weight near E$_F$ increases with decreasing temperatures, which is consistent with the expected behavior of the QP \cite{Baldassarre2008a}. In the PI phase, the temperature increase fills the gap, which is consistent with the results from Mo et al. \cite{Mo2004}. The temperature difference starting from 200 K,  $\Delta\text{T}=20$K, is the estimated temperature rise brought by the pump laser pulse for the fluence used in our pump-probe photoemission experiments (see supplementary materials). Therefore the difference curves  between high and low temperature spectra at fixed doping may serve to compare the non-equilibrium spectra with reference thermal ones.

The non-equilibrium electron dynamics has been studied with pump-probe photoemission. The differences between positive and negative time delays are shown in Fig. \ref{trPES}(\textbf{a-c}) for the PI phase. As representative of the time evolution, we consider the time-scan at -0.1 eV below E$_F$ (Fig. \ref{trPES}(\textbf{a})), whose decay can be fitted with two exponentials. The first one of 76 fs corresponds to the hot electron relaxation after photoexcitation and clearly indicates a strong electron-phonon coupling. We associate the second longer timescale of 1.7 ps with the lifetime of a transient state, as suggested by comparing the EDC's at 50 fs, 400 fs, and 2 ps with the thermal differences at equilibrium (black). At 50 fs delay (red curve) an increase in spectral weight is clearly visible both below and above E$_F$, an evidence of creation of in-gap states. The EDC cannot be fitted with a Fermi-Dirac distribution, since the system is still strongly out of equilibrium. The 400 fs delay spectrum has instead no weight above E$_F$: the excess electrons have cooled down. Nevertheless, the spectrum still deviates from the equilibrium one, in particular at -0.1 eV binding energy, indicating that, even though the electrons have relaxed, the state is different from the thermal configuration. A spectral difference equivalent to the thermal state at equilibrium can instead be found after 2 ps, when the transient state has fully relaxed. 

\begin{figure}
\includegraphics[angle=0,width=1\linewidth,clip=true]{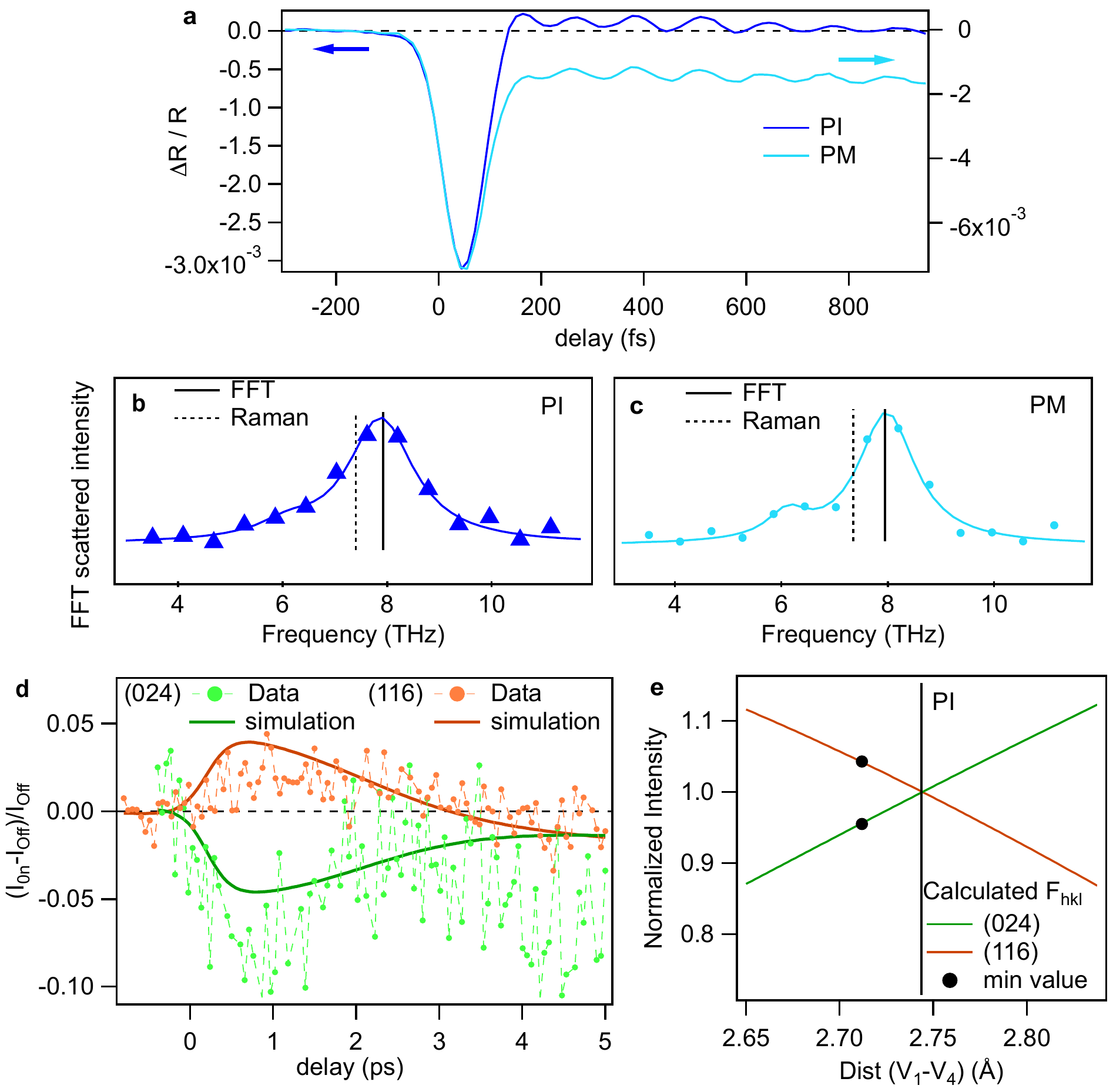}
\caption{\textbf{a)}: CPS traces for (V$_{1-x}$Cr$_x$)$_2$O$_3$ (x=0.028 PI and x=0 PM) for a fluence of 8 mJ/cm$^2$: the A$_{1g}$ coherent phonon is clearly visible \textbf{b-c)}: Fast Fourier transform of CPS traces compared to equilibrium Raman spectroscopy for the PM phase and PI phase respectively. The A$_{1g}$ pump-probe frequencies (full) present a clear blue shift compared to the equilibrium frequency (dashed) in both phases. \textbf{d)}: trXRD measurements in the PI phase for a fluence of 8 mJ/cm$^2$, showing the pump-probe diffraction peak intensities for the Bragg reflections (116) and (024). The solid lines are the simulation as explained in text. \textbf{e)} shows the calculated structure factor versus the shortest vanadium distance (V$_1$-V$_4$). The black dots represent the minimum distance observed extracted from \textbf{d}.} 
\label{hardening} 
\end{figure}

Fig. \ref{trPES}(\textbf{d-f}) reports the photoexcited behavior of pure V$_2$O$_3$ (PM) at the same fluence of 1.8 mJ.cm$^{-2}$. The time scan at 0.1 eV above E$_F$ (Fig. \ref{trPES}(\textbf{d})) shows a fast decay with a characteristic time of 70 fs and a slower one of 1.8 ps, similar to the time constants found in the PI phase. Indeed the EDC differences at 50 fs and 400 fs delays are compatible with the hot electrons not being thermalized at 50 fs and almost thermalized at 400 fs.

The observed spectral changes obtained around E$_F$ by keeping the sample at T and photoexciting with a pump pulse cannot be ascribed to heating, but rather to a genuine non-thermal transient state\cite{Wall2010,Novelli2014a}. In particular, both spectra at 50 fs and 400 fs (Fig. \ref{trPES}(\textbf{e})) suggest that there is more weight both below and above E$_F$ in the photoexcited state at temperature T than in the equilibrium state at T+$\Delta$T. Therefore, the reduction of density of states around E$_F$ is lnot compatible with a thermally excited configuration. This non-thermal state relaxes in 2 ps, similarly to the PI phase. 

Further evidence in support of a transient non-thermal phase comes from the lattice. In Fig.~\ref{hardening} we present CPS measurements that provide information on the transient response of the fully symmetric A$_{1g}$ optical phonon, which corresponds to the breathing of one entity of V$_2$O$_3$ as shown in Fig.~\ref{equi}. Consistently with previous studies\cite{Misochko1998,Mansart2010}, we observe an electronic excitation peak lasting about 200 fs, similar to the trPES response observed in Fig.~\ref{trPES}. The succeeding coherent oscillations can be analyzed by Fourier transform, which is compared in Fig.~\ref{hardening}(\textbf{b-c}) with the A$_{1g}$ mode measured with Raman spectroscopy at equilibrium. Surprisingly, the mode displays a blue-shift of up to 14$\%$ compared to the equilibrium frequency for both PI and PM phases. Such a blue-shift, i.e. a phonon hardening, is certainly non-thermal in nature. Indeed a temperature increase causes instead softening and consequently a red-shift \cite{Tatsuyama1980}. Hardening of the A$_{1g}$ phonon actually corresponds to a decrease of the average distance between the two closest vanadium atoms, d(V$_1$-V$_4$) \cite{sah09}, and as well to an overpopulation of the a$_{1g}$ orbitals (see supplementary material). It should be underlined that this coherent phonon hardening is present for both the PM and PI phases, and that its decoherence time is about 2 ps: these features are in full agreement with the behavior observed for the electronic degrees of freedom with trPES (Fig.~\ref{trPES}). There is consequently a strong evidence of a transient phase that does not correspond to any equilibrium phase of the system, involving both the electronic and lattice structure and present in both PM and PI phases.  

In order to verify our interpretation on the nature of this transient phonon blue-shift, we performed a trXRD study on the same crystals used for the trPES and CPS measurements. In Fig.~\ref{hardening}(\textbf{d-e}) we present the time dependent intensity of the Bragg reflections (116) and (204) for the PI phase. The peak positions do not change until 4 ps, therefore the lattice parameters are constant during the first few picoseconds (see supplementary material). However the intensities of both Bragg reflections vary before hand. The oxygen atoms do not affect much the diffraction intensity compared to the vanadium atoms. Supposing that the symmetry of the crystal stays the same, the diffracted intensity can be simulated by a change of the vanadium Wyckoff position, Z$_V$, and a Debye-Waller factor \cite{Johnson2009a}. The distance of the nearest vanadium atoms is given by the relation d(V$_1$-V$_4$)=(2Z$_V$-0.5)$c$, where c is the lattice constant. The (116) and (024) structure factors vary in opposite directions with Z$_V$. We find that, d(V$_1$-V$_4$) goes from 2.744 $\AA$ to a minimum value of 2.71$\AA$ before 1 ps ( d(V$_1$-V$_4$)$_{PM}$=2.69 $\AA$). The Debye-Waller is responsible for only 0.1$ \% $ of the intensity change before 4 ps. The trXRD response was not able to resolve the coherent lattice oscillations, due to limits in the signal-to-noise levels attainable during the measurements, but it does confirm that the blue-shift in the coherent phonon frequency is related to a transient reduction of the average distance d(V$_1$-V$_4$). By comparing the temporal evolution of the different experimental results, the TR-PES measurements show that the electronic structure is modified faster, and that the lattice deformation follows - which is expected for a prototype Mott system. The resulting non-thermal state is visibly more metallic in the PI phase, and seems most likely more delocalized in the PM one as well. In both cases, this state is stabilized by a transient lattice deformation that shortens the distance between the two nearest vanadium atoms and consequently increases the covalent bonding between the a$_{1g}$ orbitals. The fact that trXRD gives a slightly longer relaxation time with respect to trPES can be explained by the different probing depths of the two techniques \cite{Wegkamp2014a}.

\begin{figure}
\includegraphics[angle=0,width=1\linewidth,clip=true]{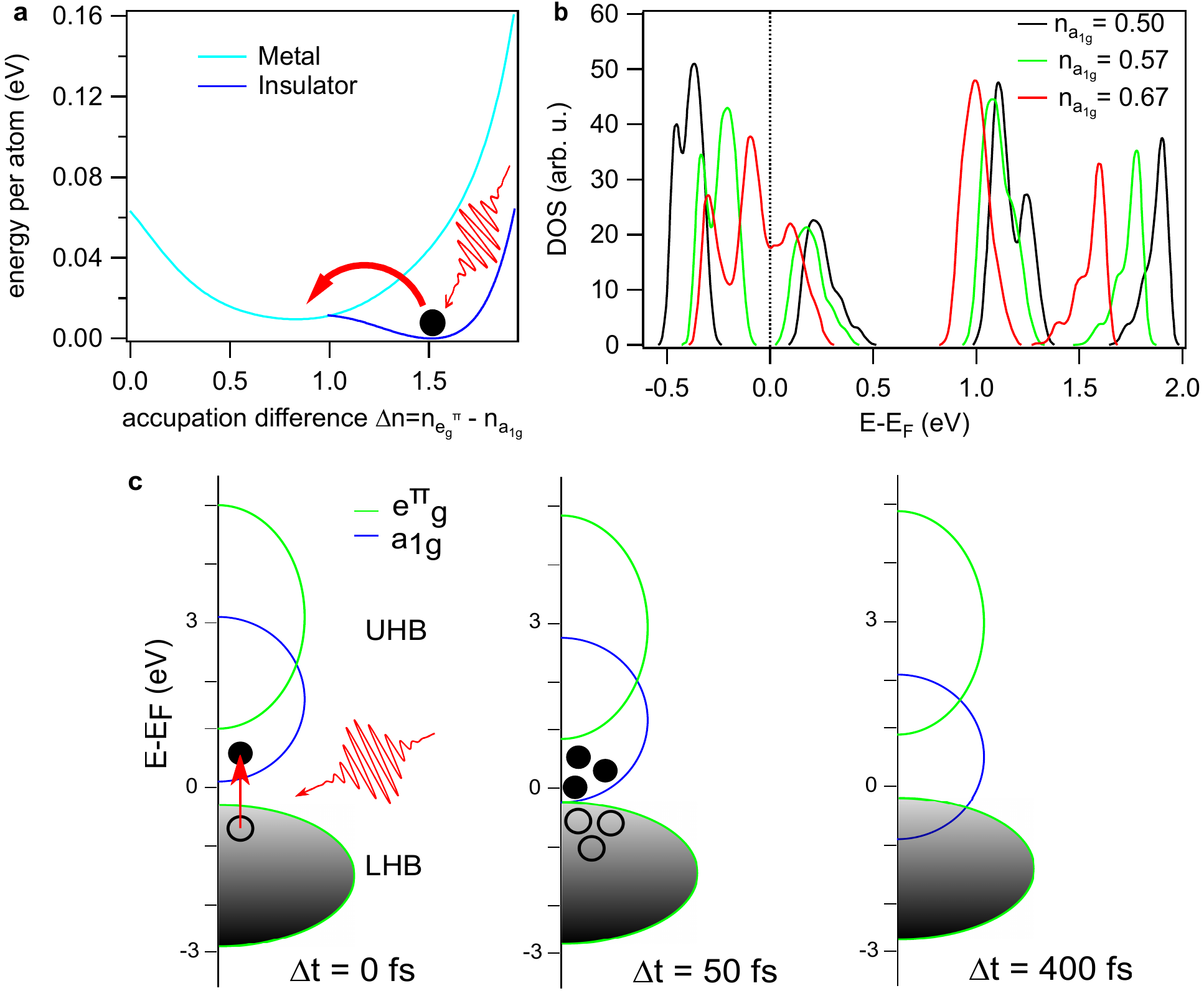}
\caption{\textbf{a)}: HF total energy as function of the occupancy difference between $e^\pi_g$ and $a_{1g}$ orbitals (the total occupancy is 2); \textbf{b)}: Density of state (DOS) for different occupancies of the a$_{1g}$; \textbf{c)} Schematic view of the proposed mechanism involved in the photoexcitation of a Mott material, where the $a_{1g}$ states lower in energy both of the PM and PI phases.} 
\label{theory} 
\end{figure}

\noindent
\textbf{Theory}:
We considered a three-band Hubbard model at one-third filling for the $t_{2g}$ orbitals with the tight-binding hopping parameters of Ref.~\cite{sah09}, and analyzed this model by means of the Hartree-Fock (HF) approximation\cite{Grieger2015} using as control parameter, after a Legendre transform, the occupancy difference between $e^\pi_g$ and $a_{1g}$ orbitals. In order to describe an insulator within an independent particle scheme as HF we had to allow for magnetism; our insulator is thus closer to the AFI low-temperature phase rather than to the high-temperature PI\cite{Grieger2015}. Within HF, the total energy, shown in Fig.~\ref{theory}(\textbf{a}), has two minima, a stable one at $\mathbf{n_{a_{1g}}}\simeq 0.5$ describes the insulator, and a metastable minimum at $\mathbf{n_{a_{1g}}}\simeq 0.625$ that instead represents a metal. In Fig.~\ref{theory}(\textbf{b}) we plot the density of states for three different values of $\mathbf{n}$, two in the insulating phase and one in the metal. We modeled the experiment in the PI phase starting from a Slater determinant that describes the HF insulator with a number of electrons transferred from the valence band of mostly $e^\pi_g$ character to the conduction one, with $a_{1g}$ character, and studied its time evolution within time-dependent Hartree-Fock.
 We find it is enough to transfer $\sim 0.13$ electrons to the conduction band to drive the system into the metastable metallic phase, as pictorially drawn in Fig.~\ref{theory}(\textbf{a}), which is consistent with the experimental excitation that are 8$\%$ for a fluence of 8 mJ/cm$^2$ in the trXrd and CPS experiments and 3.1$\%$ for the trPES . In other words, the non-thermal phase appears in this theoretical scenario as a metastable state that preexists in equilibrium and can be nucleated within the stable insulator through the photoexcitation. Since time-dependent Hartree-Fock does not account for dissipation, we cannot describe the subsequent break-up of the metastable metal nuclei back into the stable insulator, which experimentally occurs after few ps.

\textbf{Conclusion}
With a combined experimental and theoretical approach, we show that the ultrafast response of the prototype Mott-Hubbard compound (V$_{1-x}$Cr$_x$)$_2$O$_3$ is characterized by a non-thermal transient phase in which the system remains trapped before relaxing to the final thermal state.
The formation of this non-thermal phase is very fast for both PM and PI - faster than our experimental time resolution - and it is eminently electronic in nature, being driven by a transient overpopulation of a bonding $a_{1g}$ orbital. A \textit {selective} lattice deformation, strikingly highlighted by the $A_{1g}$ phonon hardening, further stabilizes this non-thermal transient phase, whose lifetime grows up to few ps: this direct comparative analysis of the evolution of the metallic and insulating phases is relevant for all the efforts aiming at photoinducing phase transitions in correlated materials, including possible technological applications like ultrafast switches.  Our results thus show that a selective electron-lattice coupling can play an important role in out-of-equilibrium Mott systems, even though the main actor remains the strong correlation; and appear to be of very general validity, suggesting that similar non adiabatic pathways can be found in other multi-band Mott compounds following excitation with ultrafast light pulses.

\textbf{Methods:} See Supplementary materials.

{\em Acknowledgments.} GL, DG, EP, MF and MM acknowledge financial support by the EU/FP7 under the contract Go Fast (Grant No. 280555).  GL, NM, LP, EP, and MM acknowledge financial support b "Investissement d'Avenir Labex PALM (ANR-10-LABX-0039-PALM), by the  Equipex ATTOLAB (ANR11-EQPX0005-ATTOLAB) and by the R\'{e}gion Ile-de-France through the program DIM OxyMORE. DB acknowledges the financial support of the French Procurement Agency (DGA) of the French Ministry of Defense. The Advanced Light Source is supported by the Director, Office of Science, Office of Basic Energy Sciences, of the U.S. Department of Energy under Contract No. DE-AC02-05CH11231. Use of the Linac Coherent Light Source (LCLS), SLAC National Accelerator Laboratory is supported by the U.S. Department of Energy, Office of Science, Office of Basic Energy Sciences under Contract No. DE-AC02-76SF00515.

\bibliographystyle{apsrev}

\end{document}